\documentclass[11pt,a4paper]{article}
\usepackage{jheppub}

\def\d{{\textrm d}}

\title{Giant Tachyons in the Landscape}
 \author[a]{Iosif Bena,} \author[a]{Mariana Gra\~na,} \author[a]{Stanislav Kuperstein} \author[b]{and Stefano Massai}

 \affiliation[a]{Institut de Physique Th\'eorique, CEA
 Saclay, CNRS URA 2306 \\ F-91191 Gif-sur-Yvette, France} 

 \affiliation[b]{Arnold Sommerfeld Center for Theoretical Physics,\\
  Theresienstr. 37, 80333 M\"unchen, Germany} 

\emailAdd{iosif.bena@cea.fr}\emailAdd{mariana.grana@cea.fr}\emailAdd{stanislav.kuperstein@cea.fr}
\emailAdd{stefano.massai@lmu.de}

\abstract{We study the dynamics of localized and fully backreacting anti-D3 branes at the tip of the Klebanov-Strassler geometry. We use a non-supersymmetric version of the Polchinski-Strassler analysis to compute the potential for anti-D3 branes to polarize into all kinds of five-brane shells in all possible directions. We find that generically there is a direction along which the brane-brane interaction is repulsive, which implies that anti-D3 branes are tachyonic. Hence, even though anti-D3 branes can polarize into five-branes, the solution will most likely be unstable. This indicates that anti-D3 brane uplift may not result in stable de Sitter vacua.}

\preprint{IPhT-T14/172, LMU-ASC 64/14}

\begin{document}
\maketitle

\setcounter{footnote}{0}
\setcounter{figure}{0}
\setcounter{equation}{0}

\section{Introduction}

Generic string theory flux compactifications with stabilized moduli
yield four-dimensional spacetimes with a negative cosmological
constant, and adding anti-D3 branes to regions of high warp factor in
these compactifications is one of the most generic methods to uplift
the cosmological constant and produce a landscape of de Sitter vacua in String Theory \cite{Kachru:2003aw}. Indeed, the prototypical example of a region with D3 brane charge dissolved in fluxes is the Klebanov-Strassler (KS) warped deformed conifold solution \cite{Klebanov:2000hb} and probe anti-D3 branes in this background have been argued by Kachru, Pearson and Verlinde (KPV)~\cite{Kachru:2002gs} to give rise to metastable configurations that describe metastable vacua of the KS gauge theory.

This intuition was challenged by the fact that the supergravity solution describing back-reacting anti-D3 branes in the Klebanov-Strassler solution must have a certain singularity in the infrared, both when the anti-D3 branes are smeared on the $S^3$ at the bottom of the deformed conifold \cite{Bena:2009xk,Bena:2011wh,Massai:2012jn,Bena:2012bk}, and also when they are localized \cite{Gautason:2013zw}. Furthermore, it was shown that this singularity cannot be cloaked with a black hole horizon \cite{Bena:2012ek,Blaback:2014tfa}, nor via polarization into D5 branes at a finite distance away from the KS tip \cite{Bena:2012vz}. Thus, all the calculations that have been done so far, which a-priori could have given either a positive or a negative or an undetermined answer about this singularity being physical, have given (via some rather nontrivial mechanisms) a negative answer.

It is important to stress that all previous works have been focused on studying properties of the anti-D3 brane supergravity solution, while the true solution which is believed to be dual to a metastable state in the KS theory is the one corresponding to anti-D3 branes polarized into NS5 branes at the tip of the KS geometry. Thus, one may argue that the infrared singularities simply signal that essential infrared physics has been ignored, as common for gravity duals of non-conformal and less supersymmetric theories. 

It is the purpose of this paper to elucidate this infrared physics. Our final result is that the anti-D3 branes can polarize into NS5 and many other types of $(p,q)$ 5-branes wrapping various two-spheres at the bottom of the KS solution. However, to our great surprise, we find that the theory describing these anti-D3 branes has a tachyonic instability which indicates that the polarized vacua will not be metastable but unstable. This in turn would imply that the de Sitter vacua obtained by uplifting with antibranes will be unstable. 

Our strategy for arriving to this result is to analyze the physics of anti-D3 branes that are localized at the North Pole of the $S^3$ at the bottom of the KS solution and to argue that these anti-D3 branes source an $AdS_5 \times S^5$ throat perturbed with RR and NS-NS three-form non-normalizable modes, dual to relevant deformations of the $\mathcal{N}=4$ SYM theory. Hence, the physics of these anti-branes can be captured by a non-supersymmetric version of the Polchinski-Strassler analysis \cite{Polchinski:2000uf}.\footnote{For earlier work on supersymmetric and non-supersymmetric relevant perturbations of $\mathcal{N}=4$ SYM in the context of five-dimensional gauged supergravity see~\cite{Girardello:1998pd,Girardello:1999bd,Pilch:2000fu}.}

At first glance, computing the appropriate relevant perturbations of this $AdS_5 \times S^5$ throat seems to be an unattainable goal, since the fully backreacted solution with localized anti-D3 branes in KS (which is a non-supersymmetric  solution that depends on more than ten functions of two variables) is impossible to obtain analytically with current technology. However, we find a way to overcome this problem by using the fully back-reacted solution with smeared anti-D3 branes we constructed in \cite{Bena:2012vz} and several key ingredients of the Polchinski-Strassler construction. First we use the potential of smeared anti-D3 branes to polarize into D5 branes wrapping the contractible $S^2$ of the deformed conifold at a finite distance away from the tip to calculate the polarization potential for localized anti-D3 branes in the same channel. Second, we decompose the self-dual part of the three-form flux near the North Pole in $(1,2)$ and $(3,0)$ components, and use this to express the
various quantities appearing in this potential in terms of fermion and boson bilinear deformations of the Lagrangian of the dual gauge theory. Third, we use these deformations to calculate the polarization potential of localized anti-D3 branes into NS5 branes wrapping a two-sphere inside the large three-sphere of the deformed conifold, as well as the potential felt by a probe anti-D3 brane in this background. We find that for generic parameters there is always some
direction along which this potential is negative, which indicates that anti-D3 branes in KS are tachyonic.

The D3-D5 polarization potential we are starting from depends on two
parameters that cannot be fixed unless one constructs the full
non-linear solution that interpolates between the infrared with anti-branes and
the Klebanov-Strassler ultraviolet (this has been performed only at linearized level in~\cite{Bena:2011wh}). However, in the final potential that we obtain the dependence on these two parameters drops
out. Hence, our result is very robust, and is independent of the
details of the gluing between the IR and UV regions.

Our result is in our opinion the definitive answer to the question of what is the fate of anti-D3 branes in the Klebanov-Strassler solution, and the physics it reveals fits perfectly with all the other results that have been obtained when studying fully-backreacted antibrane solutions. Indeed, one does not expect tachyonic brane configurations to give rise to a singularity that can be cloaked by a horizon, and this agrees with the absence of smooth negatively-charged black holes in KS, both smeared~\cite{Bena:2012ek, Buchel:2013dla} and localized~\cite{Blaback:2014tfa}. Second, an unstable brane can give rise to a supergravity solution that correctly captures the energy and expectation values of the corresponding unstable vacuum. This explains why the various calculations done using the perturbative anti-D3 brane solutions \cite{Dymarsky:2011pm, Bena:2010ze, Bena:2011hz, Bena:2011wh,Dymarsky:2013tna} yielded (rather non-trivially) the energy and VEVs one would expect from a solution with anti-branes. Third, the presence of a tachyon does not eliminate brane polarization - on the contrary, it makes it more likely along the tachyonic direction. This agrees with the fact that there exist supersymmetric and stable polarized D6-D8 configurations in $AdS$ space with negative D6 charge~\cite{Junghans:2014wda, Apruzzi:2013yva}. However, for supersymmetry breaking anti-D3 branes in flat space, the fact that the theory that describes the polarizing branes is tachyonic indicates that the polarized configurations will have either instabilities or a very low life time, and therefore they will not give rise to long-lived metastable vacua of the type needed for building cosmological models.

There are two frequent misconceptions when trying to understand the relation between our work on the supergravity backreaction of a stack of $N_{\overline{\textrm{D3}}}$ anti-D3 branes and the KPV calculation that finds that probe anti-D3 branes can polarize into long-lived metastable NS5 branes. The first is that our calculation is done in the regime of parameters when the anti-D3 branes backreact,  $g_s N_{{\overline{\textrm{D3}}}} \gg 1$, 
while the KPV calculation ignores the backreaction of the anti-D3 branes and thus it can only be valid in the opposite regime of parameters, $g_s N_{\overline{\textrm{D3}}}  \ll 1$; 
hence, since metastability is not robust under changing the parameters of the solution, one may hope  that a small number of anti-D3 branes polarized into NS5 branes
can still give rise to a metastable vacuum, which may go away as $g_s$ is increased. Nevertheless, this is not so: the KPV probe potential is derived by S-dualizing both the probe and the background, and considering the polarization of anti-D3 branes into D5 branes in the S-dual of the KS solution. However, in the KS duality frame, in order to have a polarized  anti-D3 shell with NS5 dipole charge, the mass of the  anti-D3 branes must be larger than that of the NS5 shell. Since NS5 branes have an extra factor of $g_s^{-1}$ in their tension, this only happens if $g_s N_{\overline{\textrm{D3}}}  \gg 1$, and this is precisely the regime where our supergravity analysis is valid. Our results indicate that extrapolating the results of the KPV probe calculation performed at $g_s N_{\overline{\textrm{D3}}}  \ll 1$ to describe D3-NS5 polarization in the KS solution misses essential physics.  

The other misconception is that the KPV extrapolated probe calculation only finds a metastable vacuum with NS5 brane dipole charge one when the ratio between the number of anti-D3 branes and the flux of the deformed conifold, $N_{\overline{\textrm{D3}}}/M$, is less than about $8\%$, while our calculation, as we will discuss in detail later, is valid in the regime of parameters when $N_{\overline{\textrm{D3}}} > M$. This is again a red herring, since one can do equally well a KPV calculation in which the NS5 dipole charge, $p_\textrm{NS5}$, is bigger then one and find that this calculation implies that there should exist metastable vacua for $N_{\overline{\textrm{D3}}} < 0.08 M p_\textrm{NS5}$, which is compatible with the regime in which we work and in which the tachyons are present. Hence, the extrapolated probe calculation misses the tachyonic terms in the regime where it overlaps with our calculation, and there is therefore no reason to trust it. Thus, the only regime of parameters where one can describe correctly anti-D3 branes polarized into NS5 branes in KS is the backreacted regime.
\\

This paper is organized as follows. In Section 2 we discuss the physics of the solution sourced by anti-D3 branes that are either smeared or localized at the bottom of the KS background. In Section 3 we briefly review how the Polchinski-Strassler analysis can be applied to our situation, and in Section 4 we read off the three parameters in the polarization potential of the anti-D3 branes and reconstruct the polarization potential in all possible channels. In particular we find that generically there always exists a direction along which probe anti-D3 branes are repelled, which indicates that anti-D3 branes have a tachyonic instability. In Section 5 we discuss the implications of this instability for the physics of anti-branes and present conclusions. Appendix A is devoted to a review of the non-supersymmetric Polchinski-Strassler construction. Appendix B contains the expansion of the RR and NS-NS three-form field strengths near the North Pole and the calculation of the ratio of the gaugino mass to the supersymmetric fermionic mass in the dual theory.

\section{Localized anti-D3 branes at the tip of the deformed conifold}

In this section we will describe in more detail the strategy outlined in the Introduction, to study the dynamics of localized anti-D3 branes at the tip of the Klebanov-Strassler geometry.

The Klebanov-Strassler (KS) solution~\cite{Klebanov:2000hb} is a supersymmetric warped solution based on the deformed conifold~\cite{Candelas:1989js}. This is a six-dimensional deformed cone over the five dimensional homogeneous space $T^{1,1} = (SU(2)\times SU(2))/U(1)$, which topologically is a product $S^3 \times S^2$. We will indicate by $\tau$ the radial direction of the cone. At the tip of the geometry ($\tau=0$) the $S^2$ shrinks smoothly and the $S^3$ has finite size, supporting $M$ units of RR three-form flux. The three-form fluxes of the solution combine in the complex form $G_3 = F_3 - (C_0 + i e^{-\phi}) H_3$ which is imaginary self-dual (ISD). See for example~\cite{Herzog:2001xk} for a review of the KS geometry.

\begin{figure}[t]
\centering
\includegraphics[trim = 0mm 210mm 0mm 10mm, scale=0.5]{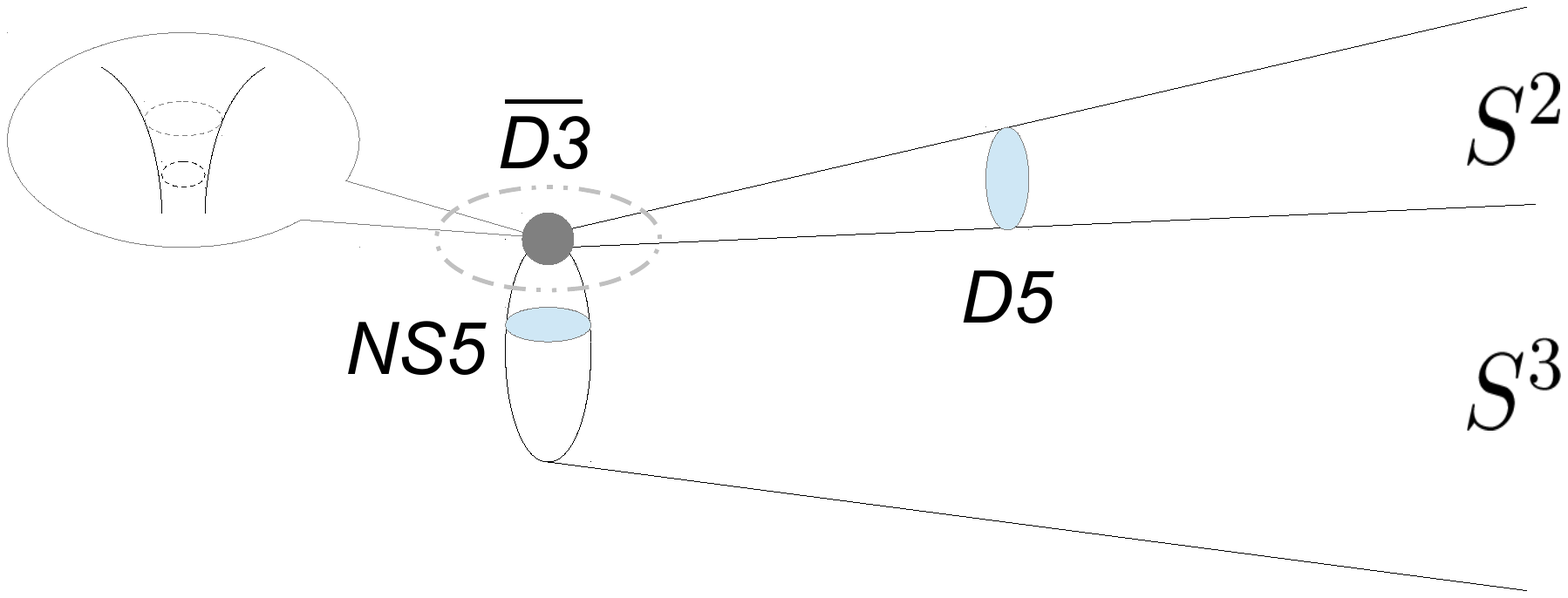}
\caption{The backreaction of localized anti-D3 branes creates an $AdS$ throat at the North Pole of the 3-sphere at the bottom of the deformed conifold. The imaginary self-dual (ISD) flux leaking into the throat becomes singular in the deep IR. We investigate the possible resolutions of this singularity by the polarization of the anti-D3 branes into D5, NS5 and other $(p,q)$ 5-branes.}
\label{fig:Localized-throat}
\end{figure}

We consider anti-D3 branes localized at one point on the large $S^3$ at the tip of this solution, which we refer to as the North Pole (NP). The deformed conifold is everywhere regular and, in particular, the vicinity of the NP locally looks like $\mathbb{R}^6$. The backreaction of anti-D3's is therefore expected to create an $AdS_5 \times S^5$ throat with a radius determined by the number of anti-branes, which we will denote throughout the paper by $N_{\overline{D3}}$. The configuration is depicted in Figure~\ref{fig:Localized-throat}. This configuration preserves one $SU(2)$ factor of the total $SU(2) \times SU(2)$ isometry group of the deformed conifold (see for example~\cite{Krishnan:2008gx,Evslin:2007ux}).

The $AdS_5 \times S^5$ throat created by the anti-D3's is glued to the ambient KS geometry, and hence it will be perturbed by modes coming from the bulk. Most of these modes will be irrelevant in the infrared, but some will not. In particular, since the anti-D3 branes preserve different supersymmetries from the KS solution, the ISD three-form flux of KS will enter the throat and create non-normalizable, relevant perturbations that correspond holographically to mass terms in the $\mathcal{N}=4$ SYM theory dual to the small throat.\footnote{A similar situation was described in~\cite{DeWolfe:2004qx} where, however, the effects of supersymmetry breaking were not taken into account.} This is precisely the situation that was considered in the work of Polchinski and Strassler (PS)~\cite{Polchinski:2000uf}. The main focus of this paper was on supersymmetric three-form flux perturbations of $AdS_5 \times S^5$, that gave the  dual of the so-called $\mathcal{N}=1^{\ast}$ theory. In our situation however the bulk perturbations will explicitly break all the supersymmetries of the anti-D3 throat, and hence we need to perform a non-supersymmetric PS analysis (similar to the one in \cite{Zamora:2000ha} and the last section of \cite{Polchinski:2000uf}). We explain this construction in all details in Appendix A. 

In the deep infrared the Polchinski-Strassler flux perturbations become important and can destroy the $AdS$ geometry, giving rise to a singularity \cite{Girardello:1999bd}. This is in line with the fact that the anti-D3 brane singularity found in the smeared and linearized solution~\cite{Bena:2009xk} is not an artefact of linearization~\cite{Massai:2012jn,Bena:2012bk} nor of the smearing~\cite{Junghans:2013xza}.\footnote{The singularity of the smeared anti-brane solution is actually milder then that of localized anti-branes: indeed the smeared solution still has anti-D3 form because the
singular fields are weaker than the fields of the smeared anti-D3 branes. However, when the anti-D3 branes are localized the singular fields become much stronger than the anti-D3 fields, and completely destroy the $AdS_5 \times S^5$ structure in the infrared~\cite{Freedman:2000xb}.} When the flux perturbations are supersymmetric, Polchinski and Strassler have shown that this singularity is resolved by the polarization of the 3-branes via the Myers dielectric effect~\cite{Myers:1999ps} into shells of $(p,q)$ five-branes, that are in one-to-one correspondence to the vacua of the dual mass-deformed ${\cal N}=4$ SYM theory \cite{Donagi:1995cf}.

Our purpose is to find whether the anti-D3 singularity can get similarly cured by the polarization of the D3 branes into $(p,q)$-five-branes with different orientations. As we explained in the Introduction, the direct route to investigate this is to solve the equations of motion to find the backreacted solution with localized supersymmetry-breaking sources, but we are not doing this. The only assumption we make is that the localized anti-D3 branes will create a flux-perturbed $AdS_5 \times S^5$ throat. Note that this assumption is minimal - if such a throat does not exist than the anti-brane solution should be disregarded as unphysical.

One of the possible polarization channels inside this throat is the one corresponding to an NS5 brane wrapping a 2-sphere inside the $S^3$ of the deformed tip, depicted in Figure \ref{fig:Localized-throat}. This channel was analyzed in the probe approximation (i.e. neglecting the backreaction of the anti-branes on the geometry) by  KPV \cite{Kachru:2002gs} and found to give rise to a locally stable configuration. Our analysis does not ignore the backreaction of the anti-D3 branes that polarize, and one of our purposes is to determine what happens to the KPV NS5 channel if one takes this backreaction into account. 

Another possible polarization channel is the ``orthogonal'' one, corresponding to D3 branes polarized into D5 branes wrapping the shrinking $S^2$ of the deformed conifold at a finite distance away from the tip (depicted also in Figure \ref{fig:Localized-throat}). As we will explain in detail below, the fact that this polarization takes place in a plane transverse to the $S^3$ allows one to compute exactly the fully backreacted polarization potential of localized anti-D3 branes by relating it to the polarization potential of smeared anti-D3 branes we computed in~\cite{Bena:2012vz}. This latter potential does not have any minima, which indicates that the effects of supersymmetry breaking are strong-enough to disable the D3$\rightarrow$D5 polarization channel of the Polchinski-Strassler analysis.
The purpose of the next section is to adapt the Polchinski-Strassler analysis to anti-D3 branes in KS and to investigate the effects of supersymmetry breaking for the NS5 polarization channel and for the oblique ones.

\section{The Polchinski-Strassler analysis of anti-brane polarization}

One of the most important results of the supersymmetric Polchinski-Strassler analysis is that the polarization potentials corresponding to different polarization channels are determined only by the UV boundary conditions that specify the relevant perturbations of the dual theory, and not by the details of the infrared geometry created by the polarized branes. Indeed, one can find the polarization potentials of the various types of branes by treating the RR and NSNS three-form field strengths dual to fermion masses as small perturbations of the original $AdS_5 \times S^5$ throat, and expanding the action of a probe five-brane in the perturbed geometry. It then turns out that, rather surprisingly, these terms are completely insensitive of the details of the infrared geometry and are solely determined by the UV boundary conditions.

Hence, to compute the polarization potentials that determine the vacua of the theory, one can simply probe the geometry sourced by un-polarized D3 branes. The potential for five-branes probing the fully backreacted polarized brane background is guaranteed to be exactly the same.\footnote{An explicit check of this can be found in~\cite{Massai:2014wba} for solutions with M2 branes polarized into M5 branes~\cite{Lin:2004nb,Bena:2004jw}.} It is very important to stress that this fact does not rely on supersymmetry, and hence it will be true also when considering relevant perturbations that break $\mathcal{N}=4$ to $\mathcal{N}=0$.

This analysis can be applied straightforwardly to antibranes localized at the North Pole of the $S^3$ in the infrared of the KS solution. We introduce complex coordinates $z_i$ for the $\mathbb{R}^6$ close to the North Pole and parameterize the location and orientation of all $SO(3)$-invariant polarized shells by a complex number $z$ such as $\mathbf{z}_i = z \cdot e^i$, where  $e_{i=1,2,3}$ is a unit real 3-vector parametrizing the $SO(3)$-rotated $S^2$ inside the $S^5$. The radius of the shell (in $\mathbb{R}^6$ coordinates) is then $\left\vert z \right\vert / \sqrt{2}$.
When supersymmetry is completely broken the five-brane probe potential depends on three parameters, $m, m^\prime$ and $\mu$. In the supergravity solution $m$ and $m^\prime$ correspond to the $(1,2)$ and $(3,0)$ components of the non-normalizable complex three-form field strength that perturb the $AdS_5\times S^5$ throat\footnote{From now on we choose the complex structure to have the same conventions as in~\cite{Polchinski:2000uf}, despite the fact that we have an anti-D3 and not a D3 throat. Hence, we will always refer to the polarizing fields as $(1,2)$ and $(3,0)$. \label{fnconvention}} \cite{Grana:2000jj,Gubser:2000vg} and the parameter $\mu$ corresponds to a certain non-normalizable harmonic scalar that transforms in the {\bf 20} of SO(6). The full polarization potential is \cite{Polchinski:2000uf,Zamora:2000ha}:
\begin{eqnarray}
\label{PolPot}
V_{(p,q)} \left( z \right) &=&  \frac{4}{\pi g_s \left( 2 \pi \alpha^\prime \right)^4} \Bigg \{ \frac{1}{N_{\overline{\textrm{D3}}}} \cdot \left\vert \mathcal{M} \right\vert^2 \left\vert z \right\vert^4  + \frac{\left( 2 \pi \alpha^\prime \right)}{3 \sqrt{2}} \textrm{Im} \left[ 3 m \overline{\mathcal{M}} z \overline{z}^2 + m^\prime \overline{\mathcal{M}} z^3 \right] \nonumber\\ 
&& \qquad + N_{\overline{\textrm{D3}}} \cdot \frac{\left( 2 \pi \alpha^\prime \right)^2}{8} \left( \left( \left\vert m \right\vert^2 + \frac{\left\vert m^\prime \right\vert^2}{3} \right) \left\vert z \right\vert^2 + \textrm{Re} \left( \mu^2 z^2 \right) \right) \Bigg\} \, , 
\end{eqnarray}
where $N_{\overline{\textrm{D3}}}$ is the number of anti-D3 branes and $\mathcal{M}$ is the mass parameter of the $(p,q)$ five-brane probe: $\mathcal{M} = p \left( C_0 + i e^{-\phi} \right) + q$.

In our solution the non-normalizable modes that specify $m$, $m'$ and $\mu$ are determined by the gluing between the region where the anti-D3 branes dominate the geometry and the asymptotically-KS UV. Since the only known solution with fully backreacted anti-D3 branes corresponds to smeared sources over the $S^3$ at the tip of the deformed conifold~\cite{Bena:2012bk}, one can try to ask what happens to the various channels of the localized anti-D3 solution when we smear the branes.

\begin{figure}[t]
\centering
\includegraphics[trim = 0mm 180mm 0mm 10mm, scale=0.25]{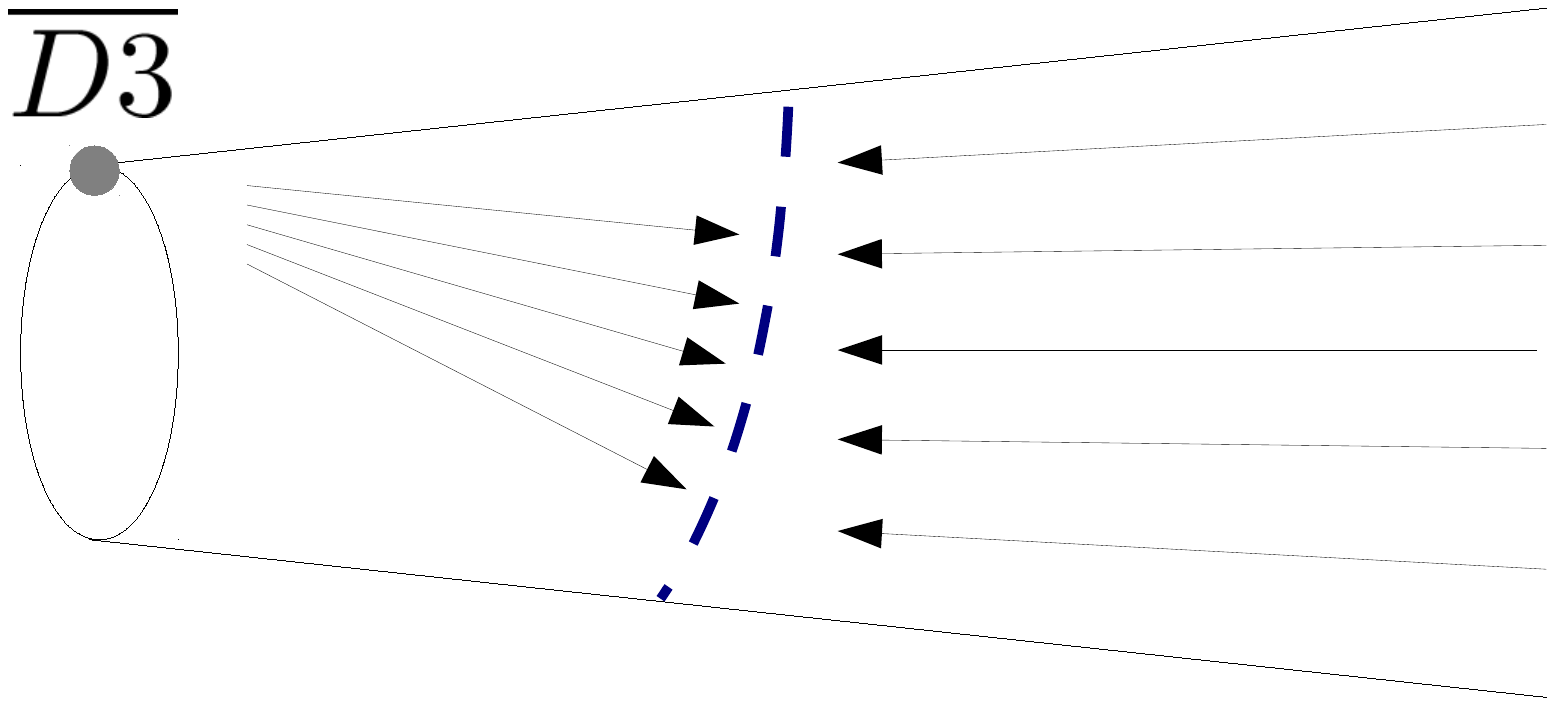}
\includegraphics[trim = 0mm 180mm 0mm 10mm, scale=0.25]{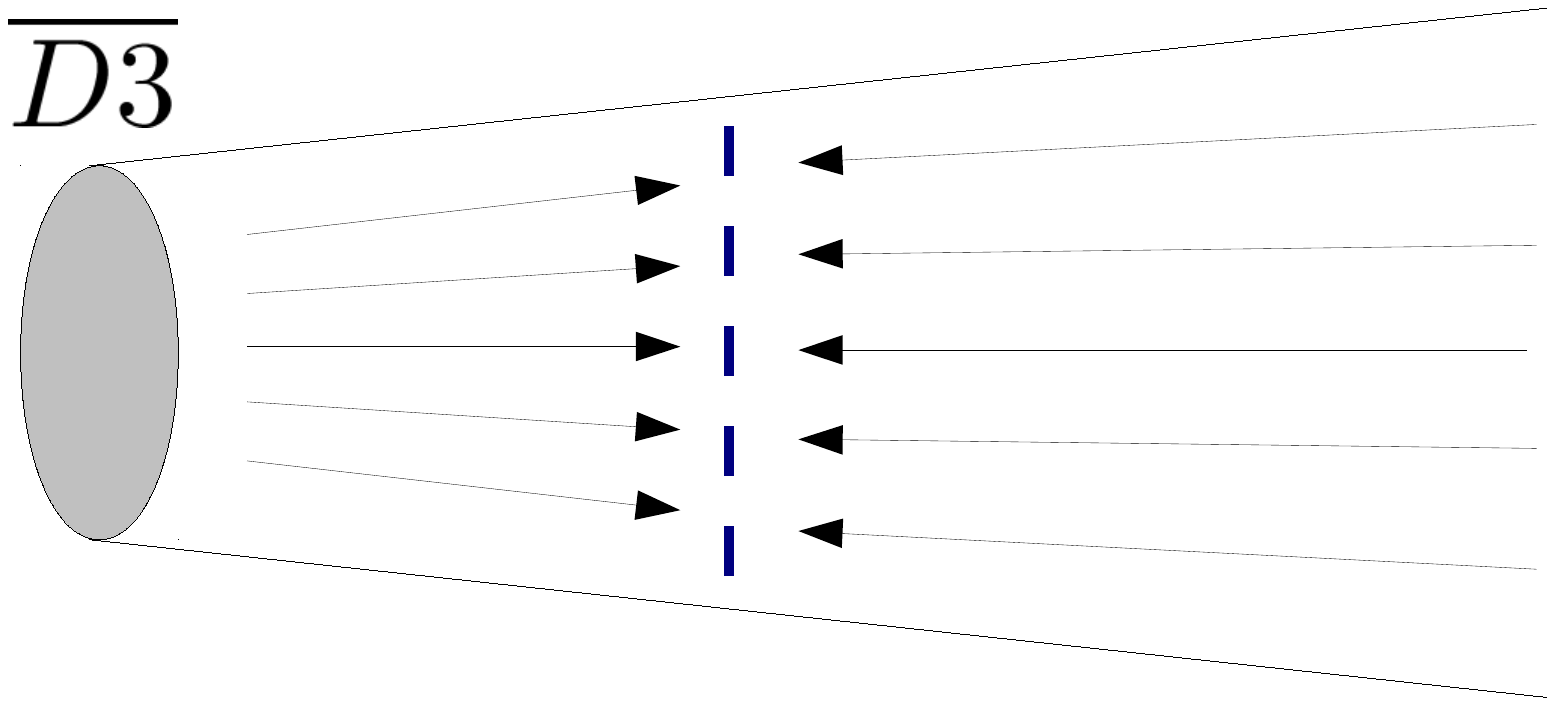}
\includegraphics[trim = 0mm 180mm 0mm 10mm, scale=0.25]{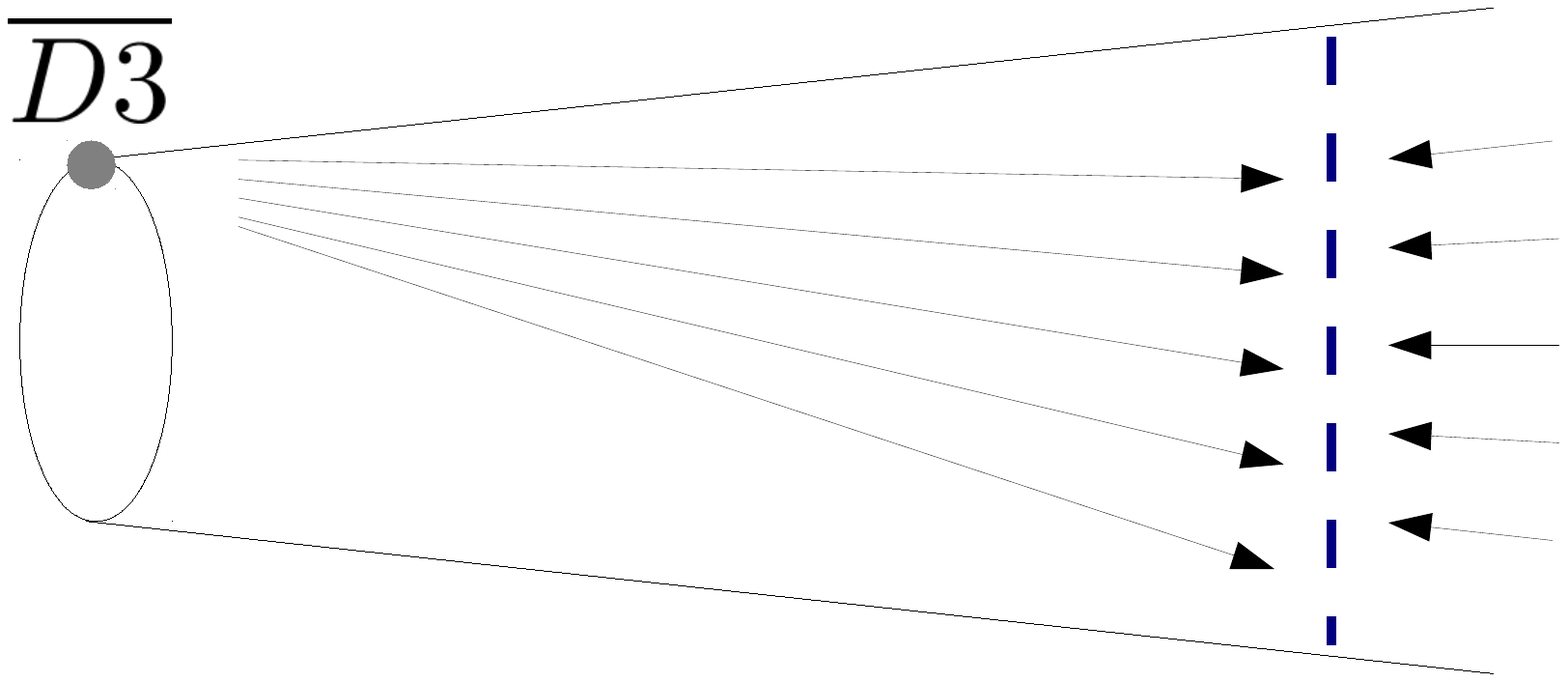}
\caption{The ``gluing surface" between the near-$\overline{D3}$ solution and the KS background is different for localized (left) and smeared sources (center). On the other hand, for a large Schwarzschild radius (right), the surface is once again $SU(2) \times SU(2)$ invariant and the mass parameters are independent of the anti-branes position at the tip.}
\label{fig:Small-Large-N}
\end{figure}

The shape of the gluing region (see Figure \ref{fig:Small-Large-N}) (which we will imprecisely refer to as ``gluing surface'') between the two regions
depends on the position of the sources and on their number. Indeed, the more anti-branes we have, the larger their Schwarzschild radius will be, and the further out the gluing surface will be pushed. Furthermore, when the anti-D3 branes are smeared on the $S^3$ at the tip of the deformed conifold, this surface corresponds to a constant radial coordinate slicing, while for a generic localized distribution of branes this surface will not respect the $SU(2)\times SU(2)$ invariance and will change its shape. Therefore, it looks like the non-normalizable modes may change when the branes are smeared or un-smeared and the shape of this surface changes. Nevertheless, we can always work in a regime of parameters where this change will be negligible: if the number of $\overline{D3}$'s is large enough, their Schwarzschild radius can be pushed away from the tip, and for $R_{\overline{D3}} \gg l_\epsilon$ the effects of moving the antibrane sources on the shape of the gluing surface and hence on the asymptotic value of the non-normalizable modes will be power-law suppressed.  

Armed with this, we can go ahead and argue that smearing of the anti-D3 branes on the $S^3$ will not affect the polarization potential for D5 branes wrapping the shrinking $S^2$ of the deformed conifold, which happens in a plane orthogonal to the $S^3$. In fact, moving the anti-D3 branes around the tip will affect the warp factor as well as $H_3$ and $F_3$. However, the cubic term in the polarization potential~\eqref{PolPot} is determined by the combination:
\begin{equation}\label{omegaplus}
  \omega_3^+ = h^{-1} (\star_6 G_3 + i G_3 ) \, ,
\end{equation}
which is both closed and co-closed ($\textrm{d} \omega_3^+ = \textrm{d} \star_6 \omega_3^+ = 0$), and therefore  it is completely determined by its asymptotic value. Hence $m$ and $m'$ do not change when the anti-D3 branes are moved. Similarly, the quadratic term has three contributions. Two of them, proportional to $m^2$ and $m'^2$ come from the backreaction of the three-forms, and are present also when the polarization is supersymmetric. Hence, they are completely determined by $m$ and $m'$ and therefore are not affected by the smearing. The third term, parameterized by $\mu$, comes from a scalar deformation that transforms in the {\bf 20} of SO(6), and since this mode is harmonic it also depends only on the data on the gluing surface. Hence, in the regime of parameters in which we are working, the polarization potential for the transverse D5 channel is not affected by the smearing of anti-D3 branes on the three-sphere.\footnote{As we will explain in Appendix A, other channels, in which the anti-D3 branes polarize into five-branes extended along the smearing direction (such as the KPV NS5 channel explored in \cite{Kachru:2002gs}) are wiped out by the smearing.}

By using this fact, we can circumvent the problem of directly computing the NS5 and oblique polarization potentials, which require the knowledge of the fully localized anti-brane solution. We will instead use the polarization potential for the D5 channel, which we computed in~\cite{Bena:2012vz} using the smeared solution, to determine the relation between $m$, $m'$ and $\mu$ and use them to reconstruct via equation~\eqref{PolPot} the potential for the NS5 and the oblique phases. 

Since we do not know the non-linear solution corresponding to smeared backreacting anti-D3 branes that interpolates between the IR and the UV (this is only known at linear level~\cite{Bena:2011wh}) our strategy is to use the most general solution sourced by anti-D3 branes compatible with the $SU(2)\times SU(2)$ symmetries of the Klebanov-Strassler background~\cite{Bena:2012bk}. This solution is parameterized by two parameters $b_f$ and $b_k$  and we will relate them to the three parameters, $m$, $m'$ and $\mu$ that enter in the polarization potential. In fact these parameters need only  be determined up to an overall scale, and we will therefore only need two relations to determine them. One such relation can be obtained directly from the transverse polarization potential computed in~\cite{Bena:2012vz}. To obtain the second relation we will use the fact that the closed ISD form decomposes into a $(3,0)$ and a $(1,2)$ 
component with respect to the conifold complex structure. From the point of view of the anti-D3 brane throat at the North Pole, the $(3,0)$ component is non-supersymmetric, and therefore corresponds in the boundary theory to the gaugino mass  $m^\prime$. Similarly, the $(1,2)$ component corresponds to the supersymmetric Polchinski-Strassler fermion mass, parameterized by $m$. Hence, the ratio of the  $(3,0)$ and $(1,2)$ components of $ \omega_3 $ near the North Pole gives the ratio $m^\prime/m$, which is enough to determine all the terms in the polarization potential.

The only condition needed to relate the D5 and the NS5 polarization potentials is that the polarization radii are sufficiently small compared to the radius of the blown-up 3-sphere. This can be done either by making the 3-sphere large enough or by increasing the D5 and NS5 dipole charge of the polarized shell, whose effect is to decrease its radius. Since the relation between the parameters $b_f$ and $b_k$ that determine the anti-D3 solution and the PS parameters $m,m'$ and $\mu$ is independent of the dipole charge, there will always exist a probe with large-enough dipole charge such that its polarization potential is unaffected by the curvature of the 3-sphere and which can therefore be used to obtain this relation. 

In the next section we will explicitly perform the computation we outlined above. A surprise awaits: the result we get is completely independent of the two integration constants $b_f$ and $b_k$ that determine the solution, and hence it does not depend at all on the gluing which determines the UV boundary conditions of the perturbed anti-D3 $AdS_5\times S^5$ throat. We will thus be able to derive a universal result regarding all polarization channels.

\section{The NS5 polarization potential and the tachyon}\label{sec:NS5polarization}

In this section we determine the polarization potential~\eqref{PolPot} for the NS5 and the oblique polarization channels. When supersymmetry is preserved ($m^\prime=\mu=0$) the orientations of the NS5 and the D5 channels correspond respectively to $z$ in (\ref{PolPot}) being purely real and purely imaginary. This is no longer true when supersymmetry is broken (an NS5 may have a lower-energy vacuum for Im($z$)$\neq 0$), but we are still interested in computing the NS5 potential for $\textrm{Im}z=0$  and the D5 potential for $\textrm{Re}z=0$, and so we will still refer to these directions as the NS5 and the D5 channels. For the two directions the $SO(3)$-invariant polarization potential (\ref{PolPot}) will be of the form:
\begin{equation}\label{VD5}
V_\textrm{NS5,D5} = a_2 \rho^2 - a_3 \rho^3 + a_4 \rho^4 \, ,
\end{equation}
where $\rho$ denotes the radius of the polarized shell ($\rho=\textrm{Re}z$ and $\rho=\textrm{Im}z$ for the NS5 and the D5 channels respectively). 
It is convenient to introduce a quantity $\lambda$ which is invariant both under the rescaling of the full potential and of the coordinate $R$:
\begin{equation}
\label{eq:lambda-def}
\lambda \equiv \frac{a_2 a_4}{a_3^2} \, .
\end{equation} 
This quantity will  be useful in relating the data of the anti-D3 brane conifold solution to the Polchinski-Strassler parameters 
$m,m^\prime$ (the fermion masses) and $\mu$ (the $L=2$ parameter).

For the D5 channel, the parameters $a_2$, $a_3$ and $a_4$ in~\eqref{VD5} were computed in~\cite{Bena:2012vz} by solving in the infrared the equations of motion for the most general ansatz compatible with the symmetries of smeared anti-D3 branes. Since the solution was not glued to the UV Klebanov-Strassler asymptotic solution, two integration constants for the flux functions, called $b_f$ and $b_k$, were not fixed.
In principle these two parameters are not independent and the precise relation between them could be determined in the UV by gluing to the KS solution or to one of its non-normalizable deformations \cite{Kuperstein:2003yt,Dymarsky:2011ve}. However, as we will see below, the physics is completely independent of the details of the UV. The potential in terms of these constants is (see (5.4) of \cite{Bena:2012vz}):
\begin{equation}\label{PotD5bfbk}
V_\textrm{D5} = \frac{\pi N_{\overline{D3}}}{3 c_0}\, \Big( b_f^2 + 48 b_k^2 \Big) \tau^2 - \frac{1}{3}b_f \,\tau^3 + \frac{c_0}{8\pi N_{\overline{D3}}}\, \tau^4 \, ,
\end{equation}
where $\tau$ is the radial coordinate of the conifold ($\tau=2 \textrm{Im} z$ near the tip) and $c_0$ is a numerical constant. The corresponding $\lambda_\textrm{D5}$ is:
\begin{equation}
\label{eq:our-lambda}
\lambda_\textrm{D5} = \frac{3}{8} \left[ 1 + 48 \left( \frac{b_k}{b_f} \right)^2\right]\, .
\end{equation}

At the same time, the value of $\lambda_\textrm{D5}$ for a general non-supersymmetric Polchinski-Strassler solution can be found  from~\eqref{PolPot} with $\textrm{Re}z=0$:
\begin{equation}
\label{eq:mu-lambda}
\lambda_\textrm{D5} = \frac{1}{4} \cdot \dfrac{m^2 + \dfrac{{m^\prime}^2}{3} - \textrm{Re}\left( \mu^2 \right)}{\left( m + \dfrac{m^\prime}{3} \right)^2} \, ,
\end{equation}
where we assumed that both $m$ and $m^\prime$ are real. Using this result we can identify $\mu^2$ (recall this is the coefficient of the $L=2$ mode) in terms of $\lambda_\textrm{D5}$, $m$ and $m^\prime$:
\begin{equation}
\label{eq:lambda-mu2}
\textrm{Re}\left( \mu^2 \right) = - 4 \left( m + \dfrac{m^\prime}{3} \right)^2 \lambda_\textrm{D5} + m^2 + \dfrac{{m^\prime}^2}{3} \, .
\end{equation}
Our purpose is to use $\lambda_\textrm{D5}$ to derive the NS5 polarization potential. Because of the extra power of $g_s^{-1}$ in the mass of the NS5 brane relative to the D5-brane, the terms in this potential have extra powers of $g_s$ which nevertheless cancel when evaluating $\lambda_\textrm{NS5}$. The only other difference between the two potentials is that the $L=2$ term has an opposite sign:
\begin{equation}
\label{eq:mu-lambdaNS}
\lambda_\textrm{NS5} = \frac{1}{4} \cdot \dfrac{m^2 + \dfrac{{m^\prime}^2}{3} + \textrm{Re}\left( \mu^2 \right)}{\left( m + \dfrac{m^\prime}{3} \right)^2} \, ,
\end{equation}
where we again used (\ref{PolPot}) for real $m$ and $m^\prime$, but this time with $\textrm{Im}z=0$. Upon substituting the value of $\mu^2$ in (\ref{eq:lambda-mu2}) this becomes:
\begin{equation}
\label{eq:lambdas-rel}
\lambda_\textrm{NS5} = \frac{1}{2} \cdot \dfrac{m^2 + \dfrac{{m^\prime}^2}{3}}{\left( m + \dfrac{m^\prime}{3} \right)^2} - \lambda_\textrm{D5} \, .
\end{equation}

In order to proceed we need to relate the parameters $b_f$ and $b_k$ appearing in~\eqref{eq:our-lambda} to the masses $m$ and $m'$. This can be done by computing the components of the three-form $ \omega_3^+$~\eqref{omegaplus} in the $AdS_5\times S^5$ throat at the North Pole, were the anti-D3 branes sit. For this we expand the deformed conifold metric around the North Pole, we choose a complex structure for the resulting $\mathbb{R}^6$ metric, and we read off the $(1,2)$  and the $(3,0)$ components of  $\omega_3^+$, which determine respectively the masses $m$ and $m'$. For simplicity, we relegate all the details of this calculation to Appendix \ref{3-form-components}, and we just state here the final result:
\begin{eqnarray}
\label{eq:(co)closed-3-form-NP}
h^{-1}&& \left(\star_6 G_3 + i G_3 \right) |_\textrm{ North~Pole} = - 4 i  \bigg[ \left( b_f + 12 b_k \right) \d \mathbf{z}_1 \wedge \d \mathbf{z}_2 \wedge \d \mathbf{z}_3  
\\
&&
+ \left( b_f - 4 b_k \right) \left( \d \mathbf{z}_1 \wedge \d \bar{\mathbf{z}}_2 \wedge \d \bar{\mathbf{z}}_3 + \d \bar{\mathbf{z}}_1 \wedge \d \mathbf{z}_2 \wedge \d \bar{\mathbf{z}}_3  + \d \bar{\mathbf{z}}_1 \wedge \d \bar{\mathbf{z}}_2 \wedge \d \mathbf{z}_3 \right) \bigg] \cdot \delta^3 + \mathcal{O} \left( \delta^4 \right)
\nonumber \, ,
\end{eqnarray}
where $\delta$ is an expansion parameter near the North Pole.\footnote{We refer to Footnote~\ref{fnconvention} for our conventions regarding the complex structure.}
One can easily check that for the NS5 ($\mathbf{x}_i=0$) and the D5 ($\mathbf{y}_i=0$) channels the $b_k$ parameter drops out from the 3-form.

The ratio of the gaugino mass to the supersymmetric mass of the other three fermions is given by the ratio of the $(3,0)$ and $(1,2)$ parts of this three-form (see equation (35) in \cite{Polchinski:2000uf} for $m_{1,2,3}=m$ and $m_4=m^\prime$).
This implies that:
\begin{eqnarray}
\label{eq:m/m}
\frac{m^\prime}{m} = \frac{b_f + 12 b_k}{b_f - 4 b_k} \, .
\end{eqnarray}
Importantly not only the ratio appears to be real but so does each of the two masses, which confirms the assumption we made in deriving equation (\ref{eq:mu-lambda}).

Armed with this knowledge we can go ahead and calculate $\lambda_\textrm{NS}$ for the potential in the NS5 channel. Plugging (\ref{eq:our-lambda}) and (\ref{eq:m/m}) into (\ref{eq:lambdas-rel}) we arrive at our first key result:
\begin{equation}
\lambda_\textrm{NS5} = 0 \, .
\end{equation}
This implies that the quadratic term in the polarization potential for this channel vanishes,\footnote{Recall that the quartic coefficient $a_4$ is always non-zero and hence $\lambda$ can vanish only when $a_2=0$.}
and hence we have:
\begin{equation}
V_\textrm{NS5} = - \frac{1}{3}b_f \, \Psi^3 + \frac{c_0}{8\pi N_{\overline{D3}}} \, \Psi^4 \, ,
\end{equation}
were now $\Psi$ is the size of the $S^2$ inside the large $S^3$ at the bottom of the deformed conifold ($\Psi \sim \textrm{Re} z$ near the NP). This conclusion may naively appear to confirm the validity of the KPV probe calculation, for which the $\Psi^2$ term also vanishes. However, our result is much deeper and more surprising. As explained in detail in Appendix A, this term in the potential represents the force felt by a mobile anti-D3 brane in the background, and its vanishing in the KPV probe calculation reflects the fact that the KS background has an $SU(2) \times SU(2)$ symmetry and therefore a single probe does not feel a force when moving on $S^3$. Our calculation, however, gives the force that a probe anti-D3 brane feels in the backreacted supergravity solution sourced by a very large number of anti-D3 branes localized at the North Pole. Since this background breaks the isometry of the three-sphere, one expects in general that probe anti-D3 branes should feel a force in this background. 
The fact that they do not, which comes after a highly non-trivial calculation, is very surprising. 
Even more surprisingly, this conclusion does not depend on the precise relation between the parameters $b_f$ and $b_k$, or in other words it is insensitive to the UV asymptotics.

This result has a very important consequence, as it implies that there exists a direction along which anti-D3 branes feel a repulsive force. To see this consider the polarization potential into $(p,q)$ five-branes wrapping a two-sphere in an oblique plane, parameterized by the phase of $z$. The fact that the quadratic term for the NS5 channel (purely real $z$) vanishes implies, from~\eqref{eq:mu-lambdaNS}:
\begin{equation}
\label{eq:Re-mu2}
\textrm{Re} \left( \mu^2 \right) =- \left(m^ 2 + \frac{{m^\prime}^2}{3}\right) \, ,
\end{equation}
 and hence the coefficient of the quadratic term along a general oblique channel is:
\begin{equation}
\label{eq:a2}
a_2 = -{C}\Big[  \textrm{Im} \left( \mu^2 \right) \textrm{Re}(z)+\textrm{Re} \left( \mu^2 \right) \textrm{Im}(z) \Big]  \textrm{Im}(z)  \, ,
\end{equation}
where $C$ is a positive constant and $\textrm{Re} \left( \mu^2 \right)$ is given in (\ref{eq:Re-mu2}).\footnote{It is trivial to check that the result vanishes for the NS5 direction ($\textrm{Im}(z)=0$) and it is positive for the D5 channel ($\textrm{Re}(z)=0$), since $\textrm{Re}(\mu^2) < 0$. Interestingly, $a_2=0$ also for $\textrm{Re} (z) / \textrm{Im}(z) = - \textrm{Re} \left( \mu^2 \right) / \textrm{Im} \left( \mu^2 \right)$.}  The crucial observation is that as long as $\textrm{Im} \left( \mu^2 \right) \neq 0$, there always exists a range of $z$ such that $a_2$ is negative:
\begin{equation}
\label{z-phase}
\frac{\textrm{Re}(z)}{\textrm{Im}(z)} = \frac{\textrm{Re} \left( \mu^2 \right)}{\textrm{Im} \left( \mu^2 \right)} \left( \gamma - 1 \right) \, ,
\end{equation}
where $\gamma$ is any real positive number.

Hence, in general there will always exist some oblique directions for which the polarization potential has a negative quadratic term. Since this term gives also the potential between unpolarized branes on the Coulomb branch, this result implies that a probe anti-brane in the $AdS_5\times S^5$ throat created by the backreacting anti-branes will be repelled towards the UV along that direction.  As we have already advertised in the Introduction, this establishes that backreacted anti-D3 branes at the tip of the KS conifold geometry have a tachyonic mode. Furthermore, this result is independent of the integration constants $b_f$ and $b_k$, which indicates that the tachyon cannot be eliminated by playing with the KS UV parameters.

\section{Conclusions and future directions}

The fact that anti-D3 branes placed in the Klebanov-Strassler geometry are tachyonic appears to be a very robust feature of their physics. Indeed, the calculation and the details of the polarization potential and the ratios of $m$ and $m^\prime$ depend  on the parameters $b_f$ and $b_k$ that determine the gluing of the Klebanov-Strassler UV with the antibrane-dominated infrared, and one might have expected on general grounds that the force between the anti-branes also depends  on these parameters. However, as we have seen, the presence of this tachyon is universal. 

This result is further supported by the presence of a tachyon \cite{Bena:2014bxa} when anti-M2 branes are added to a background with M2 brane charge dissolved in fluxes \cite{Cvetic:2000db,Bena:2010gs,Massai:2011vi}. In fact, that result appears to be stronger than the one we obtained here. The repulsion between anti-M2 branes is manifest both when they move in an oblique direction as well as on the sphere at the tip, while anti-D3 branes can move on the sphere with no force and only feel a repulsive force when moving off-diagonally. The reason behind this is that the four-sphere at the bottom of the CGLP geometry is not a four-cycle on the seven-dimensional base, and the most general anti-M2 brane solution constructed in \cite{Bena:2014bxa} allows for a change of the integral of the four-form around this four-sphere, which is not topologically protected. If one turns off this mode one finds that the potential between two anti-M2 branes is also flat along the $S^4$ at the bottom at the solution, as we found for anti-D3 branes. When one turns this mode back on, the strength of the tachyon increases by the square of the coefficient of this mode. This again confirms our intuition that the tachyon is a generic feature of the physics of anti-branes in backgrounds with charge dissolved in fluxes, that cannot be removed by playing with the parameters of the supergravity solution. It would be clearly important to confirm this explicitly by extending our analysis to other backgrounds with charge dissolved in fluxes, both with anti-D3 and with other anti-brane charges \cite{Blaback:2010sj,Blaback:2011nz,Blaback:2011pn,Blaback:2012nf,Bena:2012tx,Blaback:2014tfa}. 

An interesting future direction is to determine whether there is any way to see this tachyon by performing a KPV-like probe calculation. As we explained in the Introduction, anti-D3 branes only polarize into NS5 branes when $g_s N_{\overline{\textrm{D3}}}  \gg 1$, which is precisely the regime of parameters that our supergravity backreacted calculation captures. Nevertheless, one may consider the polarization of anti-D3 branes into D5 branes wrapping an $S^2$ inside the $S^3$ at the bottom of the solution that is obtained by S-dualizing the KS solution, and this polarization can happen in the regime of parameters $g_s N_{\overline{\textrm{D3}}}  \ll 1$, where the D5 brane DBI action used in KPV is not invalidated by large $g_s N_{\overline{\textrm{D3}}} $ effects. If our result about the tachyon is universal, this tachyon should be visible in this regime as well. Since the coefficient of the tachyon is proportional to the square of the three-form field strength, this tachyon would probably come out from terms in the brane action that are quadratic in the supergravity fields, and hence are not captured by the DBI action.
It would be very interesting to identify these terms and see whether they give rise to a tachyon. The outcome would be interesting either way: if a tachyon exists this implies that one has to reconsider many non-supersymmetric brane probe calculations done using the Born-Infeld action and see whether these calculations are invalidated by the presence of the terms that give rise to a tachyon. If a tachyon does not exist this would reveal the first instance in string theory where a tachyon goes away when changing duality frames, which would a highly unusual and hence very exciting result. 

Since our calculation is valid in the regime of parameters where the number of anti-D3 branes is large ($N_{\overline{\textrm{D3}}} > M^2$) and the gluing surface is far-away from the KS tip, one can ask whether our results will persist when the number of antibranes is smaller than $M^2$. The regime $N_{\overline{\textrm{D3}}} < M^2$ was considered in \cite{DeWolfe:2004qx}, which studied the polarization potential outside the anti-D3-dominated region and ignored the effects of the supersymmetry breaking on the quadratic term of this potential (which, as we saw in this paper, are responsible for the tachyon). To ascertain the presence of a tachyon in the regime $N_{\overline{\textrm{D3}}} < M^2$, one has to include the effects of the $SO(3)$-breaking harmonics sourced by the localized anti-D3 branes. This was done for D3 branes in \cite{Krishnan:2008gx, Pufu:2010ie}, but here the calculation will be more involved because of the broken supersymmetry. However, the robustness of the calculations done so far, that reveal the omnipresence of tachyons in anti-brane solutions, makes it unlikely in our opinion that the tachyon will go away.

Another feature that is important to understand  is what is the endpoint of the tachyonic instability. Indeed, our calculation reveals the existence of a tachyon that manifests itself by the repulsion of antibrane probes by backreacted anti-branes localized at the North Pole, but does not allow us to track what happens after the anti-branes are repelled outside of the near-North-Pole region. A similar (brane-brane repulsion) tachyon exists in $AdS_5$ solutions constructed in Type 0 string theory \cite{Klebanov:1998yya} and possibly also in supergravity solutions corresponding to non-BPS branes~\cite{Bertolini:2000jy}; if there is any relation between those tachyons and ours this would help in understanding its endpoint.

It is also important to elucidate what are the implications of this tachyon for the stability of the configurations where the anti-D3 branes polarize into NS5 branes wrapping an $S^2$ inside the $S^3$. Indeed, our tachyon does not affect the existence of a minimum in the NS5 polarization channel, and appears even to encourage brane polarization along the oblique directions in Equation~\eqref{z-phase}. However, the fact that the inter anti-D3 potential is now repulsive implies that the D3-NS5 polarized configurations will not be long lived and will most likely be unstable. Indeed, the repulsive potential makes the tunneling barrier for shooting out an anti-D3 brane from the polarized shell very shallow. Another possible effect of the tachyon is to cause a non-spherical (ellipsoidal) instability in the polarized shells \cite{Cucumber}. Hence, such a construction will not give a long-lived de Sitter vacuum, but will either give an unstable one or one whose cosmological constant will jump down whenever the anti-D3 branes are shot out. 

The fact that anti-D3 branes are unstable is also consistent with many other calculations and expectations about their physics. First, it is known that the perturbative construction of the anti-D3 brane solution~\cite{Bena:2009xk, Dymarsky:2011pm} passes some non-trivial checks~\cite{Dymarsky:2011pm, Bena:2010ze, Bena:2011hz, Bena:2011wh, Dymarsky:2013tna}. There is no conflict between this and the instability of the anti-branes. Indeed, there are many black holes and black rings that are unstable, and these solutions make perfect sense from the point of view of the $AdS$-CFT correspondence - they are dual to an unstable phase of the gauge theory, and their instability simply indicates that the dual gauge theory wants to go to a different ground state. 

This instability is also consistent with the fact that one cannot construct a black hole with anti-D3 brane charges at the bottom of the KS solution~\cite{Bena:2012ek, Buchel:2013dla, Blaback:2014tfa}: the presence of a tachyon probably makes such a black hole solution time-dependent. Presumably a similar phenomenon happens if one perturbs a black hole in $AdS_5\times S^5 $ with a dimension-two operator dual to a tachyonic deformation of the  ${\cal N}=4$ SYM gauge theory of the form $\Phi_1^2 + \Phi_2^2+\Phi_3^2-\Phi_4^2-\Phi_5^2-\Phi_6^2$, and it would be interesting to study this system in more detail. 

Last but not least, anti-branes have been used to construct solutions dual to microstates of the D1-D5-p near-extremal black hole \cite{Bena:2011fc,Bena:2012zi}, and in the probe approximation these anti-branes appear to be metastable, much like in all other anti-branes studied in this way~\cite{Kachru:2002gs,Klebanov:2010qs,Massai:2014wba}. However, on general D1-D5 CFT grounds we expect these microstate solutions to be unstable, and this instability gives the Hawking radiation rate of the dual CFT microstate \cite{Chowdhury:2008uj}. In the well-known JMaRT solution \cite{Jejjala:2005yu} this instability is visible from supergravity because the solutions have an ergo-sphere but no horizon \cite{Cardoso:2005gj} and the time scale of the instability is matched perfectly by the emission time from the dual field theory microstate \cite{Chowdhury:2007jx,Avery:2009tu}. Hence, if the instabilities of anti-branes were universal and the near-extremal microstate solutions constructed by placing negatively-charged supertubes inside BPS microstates were unstable, this would fit perfectly with what one expects from the dual D1-D5 CFT and from the general properties of non-extremal black hole microstates.

\acknowledgments{We would like to thank U. Danielsson, F. Denef, B. Freivogel, I. Klebanov, J. Maldacena, L. McAllister, M. Petrini, S. Pufu, M. Porrati and T. Van Riet for useful discussions.  
 I.B. and S.K. are supported in part by the ERC Starting Grant 240210 -- String-QCD-BH and by the John Templeton Foundation Grant 48222. I.B. is also supported by a grant from the Foundational Questions Institute (FQXi) Fund, a donor advised fund of the Silicon Valley Community Foundation on the basis of proposal FQXi-RFP3-1321 to the Foundational Questions Institute. This grant was administered by Theiss Research. The work of M.G. and S.K. is supported in part by the ERC Starting Grant 259133 -- ObservableString. The work of S.M. is supported by the ERC Advanced Grant 32004 -- Strings and Gravity.}

\appendix

\section{Review of the non-supersymmetric Polchinski-Strassler polarization}

\label{sec:PS-review}

In this Appendix we review the main aspects of the polarization of D3 branes into five-branes studied in~\cite{Polchinski:2000uf} that we use in the analysis in this paper.

The low-energy world-volume theory of a stack of $N$ (anti) D3-branes is  $\mathcal{N}=4$ Super Yang-Mills, and the 
$R$-symmetry of this theory (that rotates its six real scalars $\phi_a$, with $a=1, \ldots, 6$) corresponds to the $SO(6)$ isometry of the 5-sphere in the dual $AdS_5 \times S^5$ geometry. In $\mathcal{N}=1$ language these scalars are paired into three chiral superfields $\Phi_i=\phi_i + i \phi_{i+3}$ for $i=1,2,3$. Each of these multiplets has a Weyl spinor $\lambda_{i=1,2,3}$, which together with  $\lambda_4$, the gaugino of the vector multiplet, transform in the $\textbf{4}$ of $SU(4)$, the covering group of $SO(6)$. Giving generic masses $m_{1,2,3}$ to the three chiral multiplets leads to an $\mathcal{N}=1$ theory, with further supersymmetry enhancement to $\mathcal{N}=2$ for $m_1=m_2$ and $m_3=0$. Giving a non-zero mass $m'$ to $\lambda_4$, on the other hand, breaks supersymmetry completely, since the gaugino belongs to the vector multiplet. When all three masses are equal,  $m_1=m_2=m_3\equiv m$, the solution is $SO(3)$ invariant.

On the gravity side giving mass to the fermions corresponds to turning on non-normalizable modes of the complex 3-form flux defined as $G_3 \equiv F_3 - \left(  C_0 + i e^{-\phi} \right) H_3$.
It was first noticed by Girardello, Petrini, Porrati and Zaffaroni (GPPZ) \cite{Girardello:1999bd} that this perturbation of $AdS_5 \times S^5$ leads to a naked singularity in the infrared, caused essentially by the backreaction of the three-forms. It was realized later by Polchinski and Strassler in \cite{Polchinski:2000uf} that the singularity is resolved via the Myers effect~\cite{Myers:1999ps}, by the polarization of the D3 branes that source $AdS_5 \times S^5$ into five-branes that wrap certain 2-spheres inside the $S^5$ at a nonzero value of the $AdS_5$ radial coordinate.

The existence of these polarized branes was first ascertained  by considering probe $(c,d)$ 5-branes\footnote{In our conventions $(1,0)$ and $(0,1)$ correspond to NS5 and D5 branes respectively.  
} with D3 charge $n$ placed inside a solution sourced by $N$ D3 branes and deformed with three-form fluxes:
\begin{eqnarray}
\label{D3-like}
\textrm{d} s^2_{\textrm D3}=h^{-1/2} \textrm{d} x_\mu \textrm{d} x^\mu + h^{1/2} \left( \textrm{d} r^2 + r^2 \textrm{d} s_{S^5} \right) ,~ {\cal F}_5 = (1+\star_{10})\! \textrm{ d}h^{-1}\! \wedge \! \textrm{d}  x_0\!\wedge \! \textrm{d}  x_1\! \wedge \! \textrm{d} x_2\! \wedge \! \textrm{d} x_3 \, ,  ~~~~~
%\! \nonumber \\
\end{eqnarray}
where $h$ is the warp factor sourced by the $N$ D3 branes and $\star_{10}$ represents the Hodge dual in the full ten-dimensional metric $\textrm{d} s^2_{\textrm D3}$.

 One can define the five-brane mass parameter\begin{equation}
\label{M}
\mathcal{M} = p \left( C_0 + i e^{-\phi} \right) + q\, ,
\end{equation}
and introduce complex coordinates $z_i$ for the $\mathbb{R}^6$ transverse to the D3 branes, such that the location and orientation of all $SO(3)$-invariant polarized shells can be parameterized by a complex number $z$ such as $\mathbf{z}_i = z \cdot e^i$, where  $e_{i=1,2,3}$ is a unit real 3-vector parametrizing the $SO(3)$-rotated $S^2$ inside the $S^5$. The radius of the shell (in $\mathbb{R}^6$ coordinates) is $\left\vert z \right\vert / \sqrt{2}$.

The polarization potential of $n$ D3 branes then takes the following form (we use the conventions adopted in \cite{Zamora:2000ha}):
\begin{align}
\label{PolPotApp}
V_{(p,q)} \left( z \right) &= - \frac{S}{V} = \frac{4}{\pi g_s \left( 2 \pi \alpha^\prime \right)^4} \Bigg \{ \frac{1}{n} \cdot \left\vert \mathcal{M} \right\vert^2 \left\vert z \right\vert^4  + \frac{\left( 2 \pi \alpha^\prime \right)}{3 \sqrt{2}} \textrm{Im} \left[ 3 m \overline{\mathcal{M}} z \overline{z}^2 + m^\prime \overline{\mathcal{M}} z^3 \right] \nonumber\\ 
& \qquad + n \cdot \frac{\left( 2 \pi \alpha^\prime \right)^2}{8} \left( \left( \left\vert m \right\vert^2 + \frac{\left\vert m^\prime \right\vert^2}{3} \right) \left\vert z \right\vert^2 + \textrm{Re} \left( \mu^2 z^2 \right) \right) \Bigg\} \, .  
\end{align}
where we have omitted higher-order contributions that are subleading when $n^2 \gg g_s^2 N \left\vert \mathcal{M} \right\vert$.  
The polarization potential depends on only three complex parameters: the supersymmetric mass $m$ of the three chiral multiplets, the non-supersymmetric gaugino mass $m^\prime$ and a third parameter $\mu$ that enters in the quadratic term. This is a non-supersymmetric  $SO(3)$-invariant traceless bosonic bilinear deformation that transforms as an $L=2$ mode on the five-sphere (in the {\bf 20} of  $SO(6)$).

This polarization potential is detailed balanced. Namely, it might have a local minimum only for:
\begin{equation}
\label{Polarization-Radius}
\left\vert z \right\vert \sim m \frac{n}{\left\vert \mathcal{M} \right\vert} \alpha^\prime \, ,
\end{equation}
where the quartic, cubic and quadratic terms are of the same order, and hence none of them can be ignored. Higher-order terms in the $1/n$ expansion are subleading and can be neglected.

Let us discuss the origin of the three terms in the polarization potential:
\begin{itemize}
\item The $n^{-1} \cdot \left\vert z \right\vert^4$ term comes from the expansion of the Born-Infeld action of the 5-brane and as such is always positive. It represents the mass difference between a stack of $n$ D3 branes dissolved in a 5-brane wrapped on the $S^2$ and the same stack of D3 branes without the 5-brane. This term does not depend on the mass-deformation parameters, and its form follows from the fact that the space orthogonal to the original stack of D3 branes is locally $\mathbb{R}^6$ \cite{Polchinski:2000uf}. In particular, the D3 warp factor $h$ drops out in this term, and hence this term is independent of the location of the D3 branes that source this warp factor. Hence, this term remains the same in all D3-like geometries of the form~\eqref{D3-like}, even when $h$ is not proportional to $\left\vert z \right\vert^{-4}$ and these geometries are not $AdS_5 \times S^5$.
\item The terms cubic in $z$ come from the force exerted by the perturbation three-form field strengths on the branes. These terms are proportional to $m$ or $m^\prime$, and can be computed by plugging the 6-form potentials $C_6$ and $B_6$ (Hodge-dual to $C_2$ and $B_2$) in the Wess-Zumino action of the five-brane. As shown in \cite{Polchinski:2000uf}, when the solution has the form~\eqref{D3-like} these 6-forms are completely determined by the AISD perturbation three-form $\omega_3^-$:
\begin{equation}
\label{PS-3form-ISD}
 \omega_3^- \equiv h^{-1} \left( \star_6 G_3 - i G_3 \right)  \, ,
\end{equation}  
where $\star_6$ denotes the Hodge dual in the unwrapped six-dimensional space orthogonal to the D3 branes.\footnote{Note that the $\omega_3^-$ in the Polchinski-Strassler D3 convention used in this Appendix corresponds to the $\omega_3^+$  in~\eqref{omegaplus} that is responsible for polarizing the anti-D3 branes in KS} The  equations of motion force this form to be closed and co-closed:
\begin{equation}
\label{PS-3form-closed}
  \textrm{ d}\, \omega_3^- =   \textrm{d} \star_6 \omega_3^- = 0 \, ,
 \end{equation}  
and therefore this form is completely determined by the topology of the orthogonal space and by the UV non-normalizable modes that encode the information about the mass-deformation parameters $m$ and $m^\prime$. 
Note that this is a very powerful result: when moving the D3 branes, both the three-form  $G_3$ and the warp factor $h$ changes, but the combination of these parameters that enters in the potential of the polarized branes does not!
\item The term proportional to $n$ is proportional to the square of the fermion masses perturbation, and represents the potential felt by a probe D3 in the perturbed background. The $|m|^2$ and $|m^\prime|^2$ terms come from the backreaction of $G_3$ on the metric, dilaton and five-form, which will now exert a force on probe D3 branes. The expression of these terms was derived in \cite{Polchinski:2000uf} by using supersymmetry and in \cite{Freedman:2000xb} by a direct evaluation of the backreacted solution (the square of the three-form provides a source in the equations of motion for the trace of orthogonal metric combined with the dilaton and five-form field-strength). The term proportional to $\mu^2$ comes from a bulk non-normalizable mode that is dual to an off-diagonal traceless bilinear bosonic mass deformation of the dual theory \cite{Polchinski:2000uf}. This $L=2$ mode satisfies a Laplace (Poisson's source-free) equation on the orthogonal unwrapped space, and so its solution is also determined completely by the asymptotic boundary conditions in the UV and is independent of $h$.
\end{itemize}

Hence, the  $SO(3)$-invariant polarization potential is completely determined by the three parameters $m$, $m^\prime$ and $\mu$. The key feature of this potential, which we extensively use to derive the polarization potential of the localized anti-D3 branes, is that all the terms in this potential are determined by the UV boundary conditions and do not depend on the location of the D3 branes that source the background.

Indeed, as we have argued above, the terms proportional to $n^{-1}$ and  $n^0$, as well as the second ($L=2$) part of the term proportional to $n^1$ in (\ref{PolPot}) are independent of the warp factor $h$. The only terms that might depend on the location of the branes are therefore those proportional to $|m|^2$ and $|m^\prime|^2$. However, it is easy to see that this is not so: First, when $\mu=m^\prime=0$, the polarized configurations are supersymmetric and the potential (\ref{PolPot}) should be a perfect square. Since the first two terms of this potential are independent of the location of the branes, so should be the term proportional to $|m|^2$. Second, one can see very easily that the relative coefficient between the $|m^\prime|^2$ and the $|m|^2$ terms must be $1/3$. Indeed, the three fermions of the chiral multiplets (which have mass $m$) and the gaugino (of mass $m^\prime$) transform in the $\mathbf{4}$ of the $SU(4)$ R-symmetry of the $\mathcal{N}=4$ YM theory and  thus have to enter the potential on equal footing. we say at the beginning of the paragraph:   This argument therefore establishes that all the terms in the potential (\ref{PolPot}) are independent of the location of the D3 branes.

This observation played a key role in the analysis of \cite{Polchinski:2000uf}, as it allowed to argue that the polarization potential of several probe D3 branes in the background sourced by many coincident D3 branes is the same as the one in which the D3 branes that source the background are themselves polarized into several concentric shells. Hence, the full polarization potential, \eqref{PolPot}, is given by replacing $n$ in Eq. \eqref{PolPotApp} by the total number of three-branes, $N_{\overline{\textrm{D3}}}$.
This argument made crucial use of the fact that the main interaction between the various shells comes from the D3 branes that are dissolved in them, and ignored shell-shell interaction, which is indeed irrelevant in the limits in which the calculation was done. 

Our analysis indicates that the same observation is valid for the non-supersymmetric polarization of anti-D3 branes on the deformed conifold, and also that it is independent of the space in which the anti-D3 branes sit, as long as at leading-order the geometry is anti-D3-dominated \eqref{D3-like}. However, there are two important distinctions. First, when the space transverse to the branes has some compact directions, the polarization in the channels that are extended along these directions can be affected if the size of the polarized branes is larger than the size of these directions. Second, this argument can only be used to calculate the polarization potential along channels where the polarized brane shells do not touch each other. Hence, it cannot be used to relate the polarization potential of several probe anti-D3 branes to polarize into an NS5 brane that wraps an $S^2$ inside the $S^3$ at the bottom of the KS solution  in the background sourced by localized anti-D3 branes, to the corresponding polarization potential in the background sourced by smeared anti-D3 branes, because smearing the antibranes on the $S^3$ makes the probe shells intersect and annihilate. However, it can be used to relate the potentials for polarizing into D5 branes wrapping the contractible $S^2$ in the backgrounds of smeared and localized anti-D3 branes, because the smearing can be done without the 5-brane shells touching each other, and this is one of the key facts that enters in our analysis in Section~\ref{sec:NS5polarization}.

\begin{figure}[t]
\centering
\includegraphics[trim = 0mm 220mm 0mm 10mm, scale=0.7]{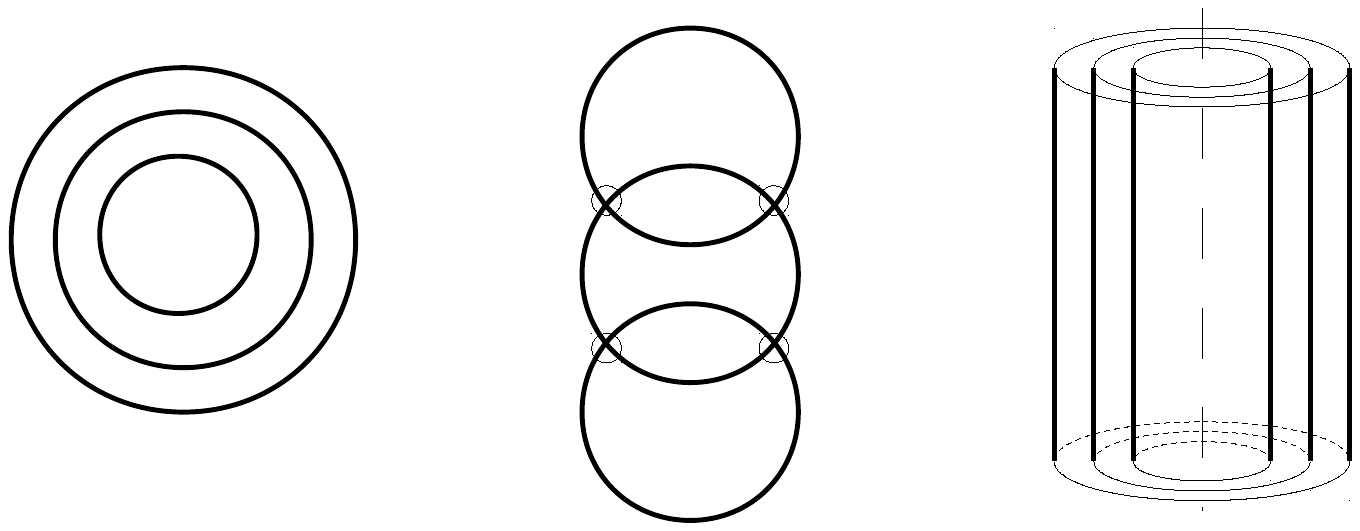}
\caption{For concentric 2-sphere shells (left) the polarization potential for a single shell is independent of the others. However when the branes are smeared inside the polarization plane and the 2-spheres intersect (center) there are new massless degrees of freedom that are not included in the DBI action and which cause the spheres to merge into cylindrical shells (right). }
\label{fig:Intersecting}
\end{figure}

\section{The complex components of the three-form at the North Pole}
\label{3-form-components}

In this Appendix we provide the details of the expansion of the three-form field strengths around the North Pole of the $S^3$ at the tip of the deformed conifold. We expand the metric around this point and we then find a complex structure for the corresponding flat six-dimensional space. We then use this parametrization to compute the components of the three-form $\omega_3^+$ defined in~\eqref{omegaplus}. From this we read off the $(1,2)$ and $(3,0)$ parts which correspond to the masses $m$ and $m'$ in the polarization potential.

\subsection*{The deformed conifold parametrization}

We follow the standard convention for the conifold (see for instance~\cite{Herzog:2001xk}). The deformed conifold is defined by $\textrm{det} W = - \epsilon^2$, where:
\begin{equation}
W \equiv \sum_{i=1}^3 z_1 \sigma^i + i z_4 \mathbf{I} = 
		\left( \begin{array}{cc}
				 z_3 + i z_4 & z_1 - i z_2 \\ 
				 z_1 + i z_2 & -z_3 + i z_4
				\end{array}  
		\right) \, .
\end{equation}
To express the $z_i$'s in terms of the angular and the radial coordinates one writes:
\begin{equation}
W = L_1 W_{(0)} L_2^\dagger
\quad \textrm{with} \quad
W_{(0)} \equiv \epsilon \left( \begin{array}{cc}
				 0 & e^{\frac{\tau}{2}} \\ 
				  e^{-\frac{\tau}{2}} & 0
				\end{array}  
		\right) \, ,
\end{equation}
where $L_1$ and $L_2$ are $SU(2)$ matrices:
\begin{equation}
L_i \equiv \epsilon \left( \begin{array}{cc}
				 \cos \frac{\theta_i}{2} e^{\frac{i}{2} \left( \psi_i + \phi_i \right)} & -\sin \frac{\theta_i}{2} e^{-\frac{i}{2} \left( \psi_i - \phi_i \right)} \\ 
				 \sin \frac{\theta_i}{2} e^{\frac{i}{2} \left( \psi_i - \phi_i \right)} & \cos \frac{\theta_i}{2} e^{-\frac{i}{2} \left( \psi_i + \phi_i \right)}
				\end{array}  
		\right) \, .
\end{equation}
This gives the following identifications:
\begin{align}
\label{eq:z1z2z3z4}
z_1 &= \frac{1}{2} \epsilon \bigg[ e^{\frac{1}{2} \left( \tau + i \psi \right)} \left( \cos \frac{\theta_1}{2} \cos \frac{\theta_2}{2} e^{\frac{i}{2} \left( \phi_1 + \phi_2 \right)} - \sin \frac{\theta_1}{2} \sin \frac{\theta_2}{2} e^{-\frac{i}{2} \left( \phi_1 + \phi_2 \right)} \right) 
\nonumber \\
&
+ e^{-\frac{1}{2} \left( \tau + i \psi \right)} \left( - \sin \frac{\theta_1}{2} \sin \frac{\theta_2}{2} e^{\frac{i}{2} \left( \phi_1 + \phi_2 \right)} + \cos \frac{\theta_1}{2} \cos \frac{\theta_2}{2} e^{-\frac{i}{2} \left( \phi_1 + \phi_2 \right)} \right) \bigg] \, ,
\nonumber \\ 
z_2 &= \frac{i}{2} \epsilon \bigg[ e^{\frac{1}{2} \left( \tau + i \psi \right)} \left( \cos \frac{\theta_1}{2} \cos \frac{\theta_2}{2} e^{\frac{i}{2} \left( \phi_1 + \phi_2 \right)} + \sin \frac{\theta_1}{2} \sin \frac{\theta_2}{2} e^{-\frac{i}{2} \left( \phi_1 + \phi_2 \right)} \right) 
\nonumber \\
&
+ e^{-\frac{1}{2} \left( \tau + i \psi \right)} \left( - \sin \frac{\theta_1}{2} \sin \frac{\theta_2}{2} e^{\frac{i}{2} \left( \phi_1 + \phi_2 \right)} - \cos \frac{\theta_1}{2} \cos \frac{\theta_2}{2} e^{-\frac{i}{2} \left( \phi_1 + \phi_2 \right)} \right) \bigg] \, ,
\nonumber \\ 
z_3 &= \frac{1}{2} \epsilon \bigg[ e^{\frac{1}{2} \left( \tau + i \psi \right)} \left( - \cos \frac{\theta_1}{2} \sin \frac{\theta_2}{2} e^{\frac{i}{2} \left( \phi_1 - \phi_2 \right)} - \sin \frac{\theta_1}{2} \cos \frac{\theta_2}{2} e^{\frac{i}{2} \left( \phi_2 - \phi_1 \right)} \right)
\\
&
+ e^{-\frac{1}{2} \left( \tau + i \psi \right)} \left( - \sin \frac{\theta_1}{2} \cos \frac{\theta_2}{2} e^{\frac{i}{2} \left( \phi_1 - \phi_2 \right)} - \cos \frac{\theta_1}{2} \sin \frac{\theta_2}{2} e^{\frac{i}{2} \left( \phi_2 - \phi_1 \right)} \right) \bigg] \, ,
\nonumber \\ 
z_4 &= \frac{i}{2} \epsilon \bigg[ e^{\frac{1}{2} \left( \tau + i \psi \right)} \left( \cos \frac{\theta_1}{2} \sin \frac{\theta_2}{2} e^{\frac{i}{2} \left( \phi_1 - \phi_2 \right)} - \sin \frac{\theta_1}{2} \cos \frac{\theta_2}{2} e^{\frac{i}{2} \left( \phi_2 - \phi_1 \right)} \right) \nonumber
\\
&
+ e^{-\frac{1}{2} \left( \tau + i \psi \right)} \left( \sin \frac{\theta_1}{2} \cos \frac{\theta_2}{2} e^{\frac{i}{2} \left( \phi_1 - \phi_2 \right)} - \cos \frac{\theta_1}{2} \sin \frac{\theta_2}{2} e^{\frac{i}{2} \left( \phi_2 - \phi_1 \right)} \right) \bigg] \, ,
\nonumber 
\end{align}
where $\psi \equiv \psi_1 + \psi_2$.

\subsection*{Expanding around the North Pole}

When written in terms of $x_i \equiv \textrm{Re} (z_i)$ and $y_i \equiv \textrm{Im} (z_i)$ the deformed conifold definition $\sum_{i=1}^4 z_i^2=\epsilon^2$ becomes:
\begin{equation}
\sum_{i=1}^{4} x_i^2 - \sum_{i=1}^{4} y_i^2 = \epsilon^2 
\qquad \textrm{and} \qquad
\sum_{i=1}^{4} x_i  y_i = 0 \, .
\end{equation}
At the North Pole of the 3-sphere we have $\left( x_4, y_4 \right) = \left( \epsilon, 0 \right)$, while the remaining six parameters $\left( x_1, x_2, x_3, y_1, y_2, y_3\right)$ provide a good set of $\mathbb{R}^6$ coordinates in the vicinity of the pole. These branes break the isometry group from $SU(2) \times SU(2)$ down to an $SU(2)$ which simultaneously rotates $\left( x_1, x_2, x_3 \right)$ and $\left( y_1, y_2, y_3 \right)$. 
In other words, we are interested in a small vicinity of the NP defined by the following $\delta$-expansion:
\begin{equation}
%\label{eq:delta-exp-1}
x_4 = \epsilon + \mathcal{O} \left(\delta^2 \right) , \,\, y_4 = \mathcal{O} \left(\delta^2 \right) 
\quad \textrm{and} \quad
x_i = \delta \cdot \mathbf{x}_i + \mathcal{O}  \left(\delta^2 \right) , \,\, y_i = \delta \cdot \mathbf{y}_i + \mathcal{O}  \left(\delta^2 \right)
\quad \textrm{for} \quad
i=1,2,3 \, .
\nonumber
\end{equation} 
We can also rewrite this $\delta$-expansion in terms of the radial coordinate $\tau$ and the angular variables appearing on the right hand side of (\ref{eq:z1z2z3z4}):
\begin{eqnarray}
\label{eq:delta-exp-2}
&&
\tau = 2 \mathbf{y} \cdot \delta + \mathcal{O}  \left(\delta^2 \right) \, , \,\,\,
\theta_1 = \frac{\pi}{2} - \alpha + v \cdot \delta +  \mathcal{O}  \left(\delta^2 \right) \, , \,\,\,
\theta_2 = -\frac{\pi}{2} - \alpha - v \cdot \delta +  \mathcal{O}  \left(\delta^2 \right) \, , \,\,\,
\nonumber \\
&&
\phi_1 = - \beta - w \cdot \delta +  \mathcal{O}  \left(\delta^2 \right) \, , \,\,\,
\phi_2 =  - \beta + w \cdot \delta +  \mathcal{O}  \left(\delta^2 \right) \, , \,\,\,
\psi =  \pi - 2 u \cdot \delta +  \mathcal{O}  \left(\delta^2 \right) \, .
\end{eqnarray}
This gives the parametrization:
\begin{eqnarray}
\begin{array}{ll}
\mathbf{x}_1 = u \cos \beta \cos \alpha - v \sin \beta
\qquad
&\mathbf{y}_1 = \mathbf{y} \cos \alpha \cos \beta
\nonumber \\
\mathbf{x}_2 = u \sin \beta \cos \alpha + v \cos \beta
\qquad
&\mathbf{y}_2 = \mathbf{y} \cos \alpha \sin \beta
\\
\mathbf{x}_3 = u \sin \alpha + w
\qquad
&\mathbf{y}_3 = \mathbf{y} \sin \alpha  \, .
\end{array} 
\end{eqnarray}

\subsection*{Expansion of the 1-forms}

The standard basis of 1-forms that diagonalize the $T^{1,1}$ metric is~\cite{Minasian:1999tt}:
\begin{align}
g_1 & = \frac{1}{\sqrt{2}}\Big( -\sin\theta_1 d \phi_1 -
  \cos \psi \sin \theta_2  d\phi_2+\sin \psi d\theta_2 \Big) \, ,
\nonumber\\
g_2 & =  \frac{1}{\sqrt{2}}\Big( d\theta_1 - \sin\psi
  \sin\theta_2 d\phi_2 -\cos\psi d\theta_2  \Big) \, ,\nonumber\\
g_3 &=  \frac{1}{\sqrt{2}}\Big( -\sin\theta_1 d \phi_1 +
  \cos\psi\sin\theta_2 d\phi_2-\sin \psi d\theta_2 \Big) \, ,\\
g_4 &= \frac{1}{\sqrt{2}}\Big( d\theta_1 + \sin\psi
  \sin \theta_2 d\phi_2 +\cos\psi d\theta_2  \Big) \, ,\nonumber\\
g_5 & = d\psi + \cos \theta_2 d\phi_2 + \cos\theta_1 d\phi_1 \, . \nonumber
\end{align}
By using the expansion~\eqref{eq:delta-exp-2} for the angles, we get to the following results for the 1-forms near the North Pole:
\begin{align}
\d \tau &= 2 \d \mathbf{y} \cdot \delta +  \mathcal{O} \left(\delta^2 \right)
\nonumber  \\
g_1 &= \sqrt{2} \left( \cos \alpha \d \beta - u \d \alpha \cdot \delta \right) +  \mathcal{O}  \left(\delta \right) 
\nonumber \\
g_2 &= - \sqrt{2} \left( \d \alpha + u \cos \alpha \d \beta \cdot \delta \right) +  \mathcal{O}  \left(\delta \right) 
\\
g_3 &= \sqrt{2} \left( u \d \alpha + v \sin \alpha \d \beta + \cos \alpha \d w \right) \cdot \delta +  \mathcal{O}  \left(\delta^2 \right) \nonumber
\\
g_4 &= \sqrt{2} \left( u \cos \alpha \d \beta + \d v \right) \cdot \delta +  \mathcal{O}  \left(\delta^2 \right) 
\nonumber \\
g_5 &= 2 \left( - \d u + v \cos \alpha \d \beta - \sin \alpha \d w \right) \cdot \delta +  \mathcal{O}  \left(\delta^2 \right) 
\nonumber
\end{align}
In terms of the local $\mathbb{R}^6$ coordinates we find:
\begin{align}
\label{expansiongR6}
\d \tau &= 2 \left( \cos \alpha \left( \cos \beta \d \mathbf{y}_1 + \sin \beta \d \mathbf{y}_2 \right) + \sin \alpha \d \mathbf{y}_3 \right) \cdot \delta +  \mathcal{O} \left(\delta^2 \right) 
\nonumber \\
g_1 &= \frac{\sqrt{2}}{\mathbf{y}} \left( -\sin \beta \d \mathbf{y}_1 + \cos \beta \d \mathbf{y}_2 \right) +  \mathcal{O}  \left(\delta \right) 
\nonumber \\
g_2 &= \frac{\sqrt{2}}{\mathbf{y}} \left( \sin \alpha \left( \cos \beta \d \mathbf{y}_1 + \sin \beta \d \mathbf{y}_2 \right) - \cos \alpha \d \mathbf{y}_3 \right) +  \mathcal{O} \left(\delta \right)
\\
g_3 &= \sqrt{2} \left( - \sin \alpha \left( \cos \beta \d \mathbf{x}_1 + \sin \beta \d \mathbf{x}_2 \right) + \cos \alpha \d \mathbf{x}_3 \right) \cdot \delta +  \mathcal{O} \left(\delta^2 \right)\nonumber
\\
g_4 &=\sqrt{2} \left( -\sin \beta \d \mathbf{x}_1 + \cos \beta \d \mathbf{x}_2 \right) \cdot \delta +  \mathcal{O}  \left(\delta^2 \right) 
\nonumber \\
g_5 &= 2 \left( - \cos \alpha \left( \cos \beta \d \mathbf{x}_1 + \sin \beta \d \mathbf{x}_2 \right) - \sin \alpha \d \mathbf{x}_3 \right) \cdot \delta +  \mathcal{O} \left(\delta^2 \right)
\nonumber
\end{align}
We note that the three $(1,0)$ forms are:
\begin{eqnarray}
g_5 - i \d \tau &=& 2 \left( - \cos \alpha \left( \cos \beta \d \mathbf{z}_1 + \sin \beta \d \mathbf{z}_2 \right) - \sin \alpha \d \mathbf{z}_3 \right) \cdot \delta +  \mathcal{O} \left(\delta^2 \right)
\nonumber \\
g_3 - i \mathbf{y} g_2 \cdot \delta &=& \sqrt{2} \left( - \sin \alpha \left( \cos \beta \d \mathbf{z}_1 + \sin \beta \d \mathbf{z}_2 \right) + \cos \alpha \d \mathbf{z}_3 \right) \cdot \delta +  \mathcal{O} \left(\delta^2 \right) 
\\
g_4 +  i \mathbf{y} g_1 \cdot \delta &=& \sqrt{2} \left( -\sin \beta \d \mathbf{z}_1 + \cos \beta \d \mathbf{z}_2 \right) \cdot \delta +  \mathcal{O}  \left(\delta^2 \right) 
\nonumber \, ,
\end{eqnarray}
where $\mathbf{z}_i \equiv \mathbf{x}_i + i \mathbf{y}_i$.

\subsection*{The closed and co-closed form}

Having obtained the explicit parametrization of the near North Pole region of the conifold, we now expand the three-form combination $\omega_3^+$ defined in~\eqref{omegaplus}, from which we can read off the mass parameters that enter in the anti-D3 polarization potential. Using the most general solution with smeared anti-D3 branes at bottom of the deformed conifold \cite{Bena:2012vz} we find the expression for $\omega_3^+$ :
\begin{align}
\label{eq:(co)closed-3-form-EXACT}
&
h^{-1} \left(\star_6 G_3 + i G_3 \right) = 4 \xi^-_k \left( - e^{-2 y} \d \tau \wedge g_3 \wedge g_4 + i g_1 \wedge g_2 \wedge g_5 \right)
\\
& \quad 
+ 4 \xi^-_f \left( - e^{2 y} \d \tau \wedge g_1 \wedge g_2 + i g_3 \wedge g_4 \wedge g_5 \right) -
2 e^{-\phi} \xi^-_F \left( -g_5 + i \d \tau\right) \wedge \left( g_1 \wedge g_3 + g_2 \wedge g_4 \right)
\nonumber \, .
\end{align}
Here $\xi_f^-$, $\xi_k^-$, $\xi_F^-$ are functions  that parameterize the fluxes and depend on the radial coordinate.  For the smeared anti-D3 solution, the functions that enter in~\eqref{eq:(co)closed-3-form-EXACT} have the following expansion for small $\tau$:
\begin{align}
&
\qquad
e^{y(\tau)} = \frac{\tau}{2} + \mathcal{O} \left(\tau^2 \right)
\, , \quad
e^{\phi(\tau)} = e^{\phi_0} + \mathcal{O} \left(\tau \right)
\, , \label{IRexp}
\\
&
\xi^-_f(\tau) = b_f + \mathcal{O} \left(\tau \right)
\, , \quad
\xi^-_k(\tau) = b_k \tau^2 + \mathcal{O} \left(\tau^3 \right)
\, , \quad
\xi^-_F(\tau) = b_F \tau + \mathcal{O} \left(\tau^2 \right)
\, .
\nonumber
\end{align}
We also know (see (4.3) of~\cite{Bena:2012vz}) that the equations of motion fix:
\begin{equation}\label{bFbk}
b_F = - 4 e^{\phi_0} b_k \, .
\end{equation}
If we now plug the expansions~\eqref{expansiongR6} as well as~\eqref{IRexp} and \eqref{bFbk}
in the expression for $\omega_3^+$~\eqref{eq:(co)closed-3-form-EXACT} we get the result~\eqref{eq:(co)closed-3-form-NP}.\footnote{In the derivation it might be useful to recall that:
\begin{equation}
\d \mathbf{x}_1 \wedge \d \mathbf{x}_2 \wedge \d \mathbf{x}_3 - i \d \mathbf{y}_1 \wedge \d \mathbf{y}_2 \wedge \d \mathbf{y}_3 = \frac{1}{4} \left( \d \mathbf{z}_1 \wedge \d \mathbf{z}_2 \wedge \d \mathbf{z}_3 + \d \mathbf{z}_1 \wedge \d \bar{\mathbf{z}}_2 \wedge \d \bar{\mathbf{z}}_3 + \d \bar{\mathbf{z}}_1 \wedge \d \mathbf{z}_2 \wedge \d \bar{\mathbf{z}}_3  + \d \bar{\mathbf{z}}_1 \wedge \d \bar{\mathbf{z}}_2 \wedge \d \mathbf{z}_3 \right) \, .
\nonumber
\end{equation}}

\bibliographystyle{utphys}
\bibliography{Final-D3}

\end{document}